# The Cardiac Analytics and Innovation (CardiacAI) Data Repository: An Australian data resource for translational cardiovascular research

Victoria Blake[1,2*], Louisa Jorm[2], Jennifer Yu[1,3], Astin Lee[4,5], Blanca Gallego[2**], Sze-Yuan Ooi[1,3**]

[1]Eastern Heart Clinic, Prince of Wales Hospital, Sydney, NSW, Australia

[2]Centre for Big Data Research in Health, University of New South Wales, Sydney, Australia

[3]Prince of Wales Clinical School, University of New South Wales, Sydney, NSW, Australia

[4]School of Medicine, Faculty of Science Medicine and Health, University of Wollongong, NSW, Australia

[5]The Wollongong Hospital, Wollongong, NSW, Australia

*Corresponding Author: Victoria Blake v.blake@unsw.edu.au

**Author Notes: Blanca Gallego and Sze-Yuan Ooi are co-senior authors of this paper. Blanca Gallego is the senior technical author and Sze-Yuan Ooi is the senior clinical author.

## Background:

Vast amounts of electronic clinical data are generated every day by healthcare providers spanning acute, primary, and community healthcare settings[1]. In addition, the proliferation of personal technology is generating previously unavailable sources of health information from smartphone health applications and wearable devices such as smartwatches[2]. Creating comprehensive datasets that capture the entire patient journey is a shared ambition amongst health researchers, governments, and clinicians with numerous potential applications for improving healthcare and generating novel insights[3].

Linked, longitudinal patient-level data is essential for adoption of a learning healthcare system that integrates research into clinical practice by continuously evaluating and improving care[3-6]. Recent advances in key technologies such as cloud computing, machine learning and natural language processing are enabling data to be accessed, linked, and analysed in real-time. Collaboration between data scientists, clinicians, policymakers and patients will ensure this cycle of innovation and evaluation can identify unmet needs improve care efficiency and quality, and rapidly assess innovations to mitigate risks and barriers to implementation[6].

Chronic diseases, such as cardiovascular disease (CVD), are an ideal focus for this integration of research into clinical pratice[3]. CVD is the leading cause of death and hospital expenditure in Australia, and significantly impacts the population, being the leading cause of disease burden by daily adjusted life years (DALY)[7-9]. Even small improvements in the efficiency of care and patient outcomes can have a significant societal and financial impact[10]. A significant barrier to the efficient delivery of cardiovascular care is the fragmented nature of the healthcare system, where multiple providers deliver care across disciplines and settings[3,10]. Clinicians are increasingly required to gather and interpret large amounts of information from multiple sources to provide consistent and quality care whilst also under pressure to increase efficiency[11]. Prospectively linking health data sources together will facilitate multi-disciplinary collaboration and the development of analytical tools that assist clinicians in managing CVD more efficiently. This, in turn, will drive the delivery of quality, evidence-based and patient-centred care[10,12,13].

Realising this goal is a significant challenge. Australian healthcare data are siloed and rarely linked, unlike nationalised healthcare systems such as in the UK and Scandinavia. Australia has a complex environment of public and private healthcare providers, which lack a unified electronic medical record (EMR) system[10,14]. There are considerable regulatory, data security and data governance requirements to satisfy when performing such large-scale data linkage across healthcare providers[3,15]. Data custodians and ethics committees are understandably cautious about approving such data linkages since they are responsible for ensuring that patient privacy is protected, research and quality improvement activities have appropriate oversight, and data is used for the benefit of patients[3,16]. Furthermore, establishing a secure environment for the data requires significant resources and funding that are not always available to small research teams.

There are also considerable technical challenges in linking health data. EMR systems often lack the ability to connect and exchange information with other systems and require proprietary software to extract information from them[17]. Information from a single hospital admission can be stored in many separate databases with limited sharing between these systems. Additionally, EMR systems were not designed for research, and so similar data types are not stored in a standardised format across these systems[15]. There are also practical challenges with using the data. Health data include a variety of different data types such as free-text, images and numerical measurements which require specialist skills, tools, and expertise to utilise them effectively[18]. Even when these technical and governance obstacles are overcome, data and data extraction processes are rarely shared, leading to inefficient and wasteful duplication by research groups seeking to extract similar information[19].

### The CardiacAI Data Repository

The Cardiac Analytics and Innovation (CardiacAI) Data Repository is an independent, not-for-profit platform that provides a secure and appropriately governed environment for Australian research groups to access and analyse real-time cardiovascular EMR data. The project was initiated by a group of senior cardiologists from South Eastern Sydney and Illawarra Shoalhaven Local Health District (LHD) and data scientists from the University of New South Wales (UNSW) Centre for Big Data Research in Health (CBDRH) with the objective of building a digital healthcare and research ecosystem for cardiovascular care. By providing the required research infrastructure and fostering strong and mutually beneficial partnerships with data contributors, the project aims to support innovative translational research that improves cardiovascular healthcare delivery and outcomes.

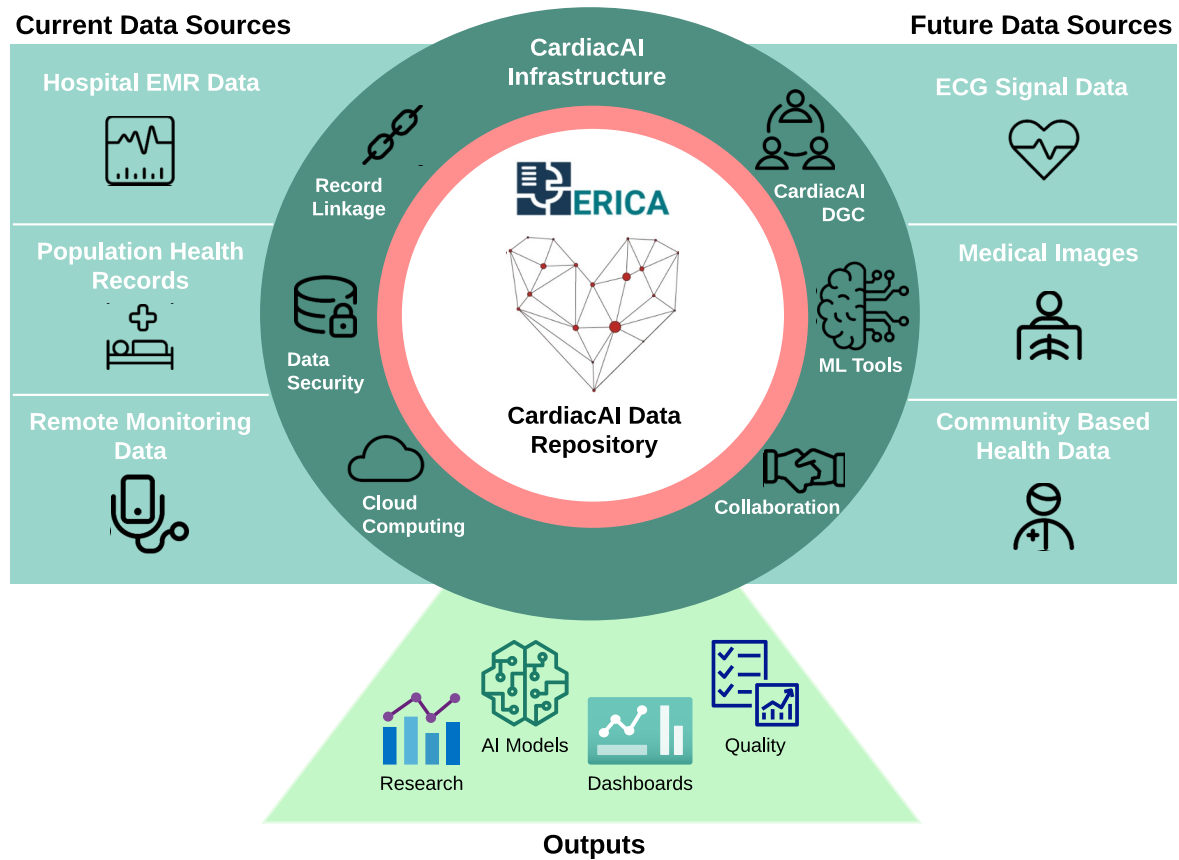

*Figure 1. The CardiacAI project*

The CardiacAI project prospectively links cardiovascular EMR and other health data sources and securely stores it in a cloud-hosted secure environment (E-Research Institutional Cloud Architecture, ERICA) designed for storing and using sensitive big health data. The project currently collects data from South Eastern Sydney Local Health District (SESLHD) and Illawarra Shoalhaven Local Health District (ISLHD), two of the 15 LHDs in New South Wales (NSW), Australia. Expansion of the data collection to include two additional LHD sites, and a set of cardiovascular telehealth datasets is underway.

To extract hospital EMR data, the team have developed custom database extraction scripts that can be easily shared with other Australian hospitals that have a Cerner Millennium[20] EMR system, the predominant system used in Australian public hospitals[21].

Extraction methods for other EMR and health data systems will be developed as required to iteratively add data sources to the project.

The cohort comprises all individuals admitted to hospital under a cardiovascular specialist at four tertiary referral hospitals, and it is defined by the following criteria:

1) Admission to Prince of Wales Hospital, St George Public Hospital, The Sutherland Hospital or The Wollongong Hospital
2) Admission under a cardiologist, cardiothoracic surgeon, or vascular surgeon
3) Admission date on or after the 1st of January 2017

EMR records meeting these criteria are extracted nightly by extraction scripts. The scripts obtain de-identified EMR data including both structured information such as coded data fields and numerical values, and unstructured data such as free-text clinical documents.

The complex, encoded EMR data are automatically transformed into a useable format for research composed of standard tables (such as the patient, encounter, diagnosis and medication administration tables), with well-defined primary and foreign keys, and with data elements using standard ontologies (e.g., SNOMED CT) when required.

These EMR records are linked with other health datasets using a technique called probabilistic linkage, whereby records are matched based on the probability that they refer to the same individual. The NSW Centre for Health Record Linkage (CHeReL) performs linkage with state-wide population health datasets annually, including the Admitted Patient Data Collection (APDC), Emergency Department Data Collection (EDDC), NSW Registry of Birth, Deaths and Marriages (RBDM) and the Cause of Death Unit Record File (COD-URF)[22]. Links to data dictionaries for these datasets can be found in the supplementary material.

These administrative data sources allow ascertainment of longitudinal outcomes for patients who are included in CardiacAI, including hospital stays, emergency department visits and mortality.

### CardiacAI ethical and governance framework and data sharing

Ethical approval for the project has been granted by the SESLHD Human Research Ethics Committee (HREC) (2019/ETH12625) and the Population Health Services Research Ethics Committee (2020/ETH01614). The project operates under an opt-out consent model whereby patients are informed about the project via information posters hung in cardiovascular wards and emergency departments. Data within the repository and research activities conducted with these data are governed by the CardiacAI Data Governance Committee (CardiacAI-DGC). This committee is comprised of the health data scientists, senior cardiovascular clinicians, and data custodians. The committee is responsible for assessing data access applications, guaranteeing the integrity and security of the data repository, and ensuring the project activities are conducted in accordance with the ethically approved project protocol, data governance framework and individual agreements with the sites contributing data. This model allows data contributors and data custodians to maintain control of the data, protect patients' and clinicians' privacy and guide the use of data in meaningful and relevant ways that will positively impact cardiovascular care.

Access to data is provided on a cost-recovery basis only, thereby avoiding duplication of the lengthy data extraction and linkage processes conducted by the CardiacAI research team, while allowing for the sustainability and improvement of the data repository. To ensure that data access is aligned with the project's primary objectives and ethical standards, researchers seeking to use the CardiacAI Data Repository must submit a project proposal to the CardiacAI-DGC for review. Proposals are assessed in accordance with the five safes framework[23] and must align with the CardiacAI project protocol aims, or be suitable for an amendment which will be evaluated by the relevant HREC.

Researchers can only access the data through UNSW ERICA[24], a highly secure, cloud platform managed by UNSW accredited by eHealth NSW under their Privacy and Security Assessment Framework (PSAF). ERICA has been built to meet the requirements of the most security-sensitive organisations with strict access controls and oversight of the movement of data in and out of the platform. More information on the data security measures employed by ERICA can be found in the [supplementary material](supplementary material). Exporting of raw data is not permitted and all research outputs are checked by senior project researchers for compliance with the project's ethical approval and adherence to statistical disclosure control principles[25].

### CardiacAI Data

A wide range of de-identified EMR and population health data are available in the CardiacAI data repository. These include diagnoses and surgical procedures, de-identified, free-text clinical documents, pathology and medical imaging investigations, physiological observations, medication and blood product administrations, dates and times of admission and ward movements. A full data dictionary can be found at [www.CardiacAI.org/data-dictionary](www.CardiacAI.org/data-dictionary) and more information describing the data can be found in the [supplementary material](supplementary material).

The clinical text is a particularly valuable source of information as it contains important clinical details that are not available in structured formats[26]. Due to the sensitive nature of this information, the CardiacAI project has taken measures to ensure the protection of patients' and clinicians' privacy. While structured personal identifiers are removed from the data extract before it is exported to CardiacAI, patient and clinician names and other personal identifiers may still be embedded in the free-text. To address this issue, the CardiacAI project has established a process that automatically removes personal identifiers from the text using a custom-built, de-identification tool developed on Australian clinical text[27] that performs with high accuracy (F1 score 90.6 to 97.17). In addition, the de-identified clinical text is stored separately from the structured data and its access is strictly controlled. Researchers must seek permission from the CardiacAI-DGC to access the clinical text and provide a compelling rationale for doing so. This approach reduces the number of people with access to the raw text, thereby reducing the risk of re-identification, while still facilitating the conduct of research.

## CardiacAI Cohort Characteristics

### Methods

The following tables have been provided to give an overview of the data within the CardiacAI data repository. The denominator for categorical characteristics displayed in the tables is the total number of encounters for that cohort. The results of our analysis have been compared with similar studies that provide demographic, comorbidity, and outcomes for each main disease group. In selecting these studies for comparison, Australian or multi-site studies were favoured over international studies or smaller, single-site studies.

### Cohort

All hospital EMR admission records (encounters) in the CardiacAI Data Repository from 1st January 2017 to 31st December 2020 were extracted as complete linked population health datasets were available for this period, thus allowing us to include outcomes, diagnosis, and procedure code in the analysis. We expect the next data linkage with updated data to be received in 2023. The encounters were linked with their corresponding records from the APDC dataset via hierarchical matching of admission and discharge date and times and facility names. From the 50,339 encounters, 94 were removed as they were missing an associated encounter record in the APDC, two were removed as they were duplicated EMR encounter records, and two were removed as they had an encounter type of "organ procurement". Figure 2 shows the encounter selection and exclusions.

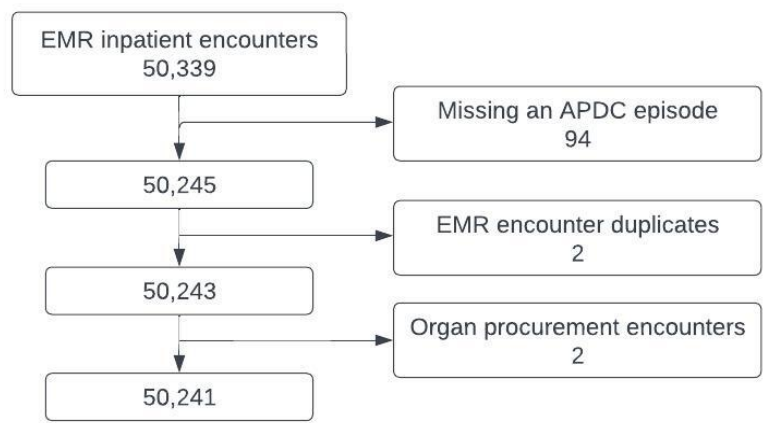

*Figure 2. EMR encounter selection diagram.*

Cardiovascular Subgroup Analysis

Four cardiovascular subgroups were selected to investigate the characteristics of common CVD patient groups. These groups were identified by the primary discharge diagnosis code assigned to that encounter. The groups and their associated International Classification of Diseases Tenth Revision Australian Modification (ICD-10-AM) codes include acute coronary syndrome (ACS) I20-25, heart failure (HF) I50, atrial fibrillation (AF) I48 and syncope R55. The subgroups were categorised at the encounter level; therefore, an individual patient may appear in more than one group if that patient had encounters that fit multiple subgroup definitions. The top primary discharge diagnosis and procedure codes for encounters that did not fit into these four groups can be found in the Table 3 of the [supplementary material](supplementary material).

Outcomes

Dates of death were derived from the RBDM and EMR dates of death, and APDC and EDDC death separation types. EMR, APDC and EDDC were used when RBDM data was unavailable. Since cause of death information was only available up to 31$^{st}$ December 2019, we have reported all-cause mortality instead of cardiovascular death. We expect to receive updated cause of death data in 2023.

It is common for patients to be transferred between facilities as a continuation of care for the same event, hence readmissions were derived by grouping the APDC episodes into 'episode sets' which were defined as a sequential set of encounters with the following criteria: 1) the admission date occurred on the same day or overnight from the previous discharge date; and 2) The mode of separation from the previous encounter was coded as a 'transfer' or 'type change separation'; or 3) The new encounter's facility name matches the facility name recorded as the location that the person was transferred to in the previous encounter.

The number of days till readmission from the index EMR encounter was calculated from the discharge date of the last encounter in the index episode set, to the first admission date of the next relevant episode set. An emergency status of 'emergency' in the first encounter in an episode set was used to determine if the readmission was unplanned. A readmission was classified as all-cause, or having a cardiovascular primary diagnosis, or cardiovascular primary or secondary diagnosis, using ICD-10-AM and Australian

Classification of Health Intervention (ACHI) diagnosis and procedure codes for major atherosclerotic, arteriovenous thromboembolic and related cardiovascular diseases[28]. The diagnosis and procedure codes from the first encounter in the readmission's episode set were used to determine its cardiovascular status. The following composite outcome categories were defined: 1) all-cause, unplanned readmission or all-cause mortality (ACread|ACmort), 2) unplanned readmission with a cardiovascular primary or secondary diagnosis or procedure or all-cause mortality (CVread|ACmort) and 3) unplanned readmission with a cardiovascular primary diagnosis or procedure or all-cause mortality (PCVread|ACmort). All post-discharge mortality figures exclude in-hospital death.

## Results with Comparison to Prior Research

### CardiacAI Cohort Characteristics

The CardiacAI cohort includes 50,241 encounters (by 36,861 patients) between 1st January 2017 and 31st December 2020. A summary of the cohort characteristics can be found in Table 1. There were no missing values in the fields that were analysed apart from language which had only three missing values. Median age was 71 and most patients (81.2%) were over the age of 55 years on admission. 60.5% were male. Aboriginal and Torres Strait Islander status was reported for 1.8% of encounters, which was lower than the reported state-wide proportion for cardiovascular hospital admissions (2.7%)[29]. For the majority of encounters, the patient was recorded as being born in Australia (59.2%) and had English as their primary language (86%). Three quarters of encounters (76.5%) were under a Cardiologist. Encounter numbers stayed stable in the three pre-COVID-19 years (2017-2019). However, 2020 saw a drop of around 700 encounters (5.5%) from the average number of encounters over the pre-COVID-19 years. This pattern is consistent with previously reported drops in cardiac encounters from the beginning of the COVID-19 pandemic and could be attributed to the suspension of non-urgent elective procedures and delayed hospital presentations due to fear of contracting COVID-19[30-32].

The majority (65.1%) of encounters were classed as emergency (non-planned) admissions and the average length of stay (LOS) was 4.2 days (+/- 6.5 days). The average LOS was slightly lower than the average LOS for patients with CVD in Australia (5.9 days) as reported by the Australian Institute of Health and Welfare (AIHW)[8]. However, this could be attributed to the way the CardiacAI cohort is defined. CardiacAI includes all patients admitted under a vascular surgeon and therefore the dataset includes non-cardiovascular procedures such as varicose vein procedures with associated shorter hospital stays. In addition, CardiacAI does not currently include stroke patients who generally have a longer length of stay[8]. The most common primary diagnosis code was I48.9 - "Atrial fibrillation and atrial flutter, unspecified" (8.7%) and the most common primary procedure code was 38218-00 - "Coronary angiography with left heart catheterisation" (10.5%).

Less than one in 10 patients (8.6%) were admitted to an Intensive Care Unit (ICU), and in-hospital mortality occurred in 1.5% of encounters. Unplanned readmission or mortality rates at six months post-discharge were 31.6% for ACread|ACmort, 18.2% for CVread|ACmort and 13.8% for PCVread|ACmort. Most unplanned readmissions (87%) occurred in SESLHD and ISLHD hospitals. However, CardiacAI only captured EMR data for around a quarter (27%) of these readmissions. This discrepancy may be explained by readmissions under other subspecialties or to health facilities outside of SESLHD and ISLHD. This highlights the value of the linked data to capture readmission events from other specialties, hospitals from other LHDs or private facilities.

*Table 1 CardiacAI cohort characteristics*

| Measure | | Result |
|---|---|---|
| Patients | | 36861 |
| Encounters | | 50241 |
| Speciality | Cardiology | 38433 (76.5%) |
| | Vascular Surgery | 8304 (16.5%) |
| | Cardiothoracic Surgery | 3504 (7.0%) |
| Admission Year | 2018 | 12802 (25.5%) |
| | 2019 | 12795 (25.5%) |
| | 2017 | 12612 (25.1%) |
| | 2020 | 12032 (23.9%) |
| Sex | Male | 22298 (60.5%) |
| Age | Mean (SD) | 68.5 (15.3) |
| | Median [Min, Max] | 71.0 [14.0,106.0] |
| | <45 | 4097 (8.2%) |
| | 45-54 | 5332 (10.6%) |
| | 55-64 | 9285 (18.5%) |
| | 65-74 | 12921 (25.7%) |
| | >=75 | 18606 (37.0%) |
| Indigenous Status | ABTSI | 657 (1.8%) |
| Country of Birth | Australia | 21829 (59.2%) |
| Language | NESB | 7020 (14.0%) |
| Length of Stay | Mean (SD) | 4.2 (6.5) |
| | Median [Min, Max] | 2.1 [0.0,293.5] |
| | <= 24 hrs | 13130 (26.1%) |
| | 1-3 days | 17155 (34.1%) |
| | 3-7 days | 11118 (22.1%) |
| | 7-14 days | 5961 (11.9%) |
| | >14 days | 2877 (5.7%) |
| Admission Type | Emergency | 32688 (65.1%) |
| Outcomes - In-Hospital | ICU Admission | 4299 (8.6%) |
| | All-Cause Mortality | 740 (1.5%) |
| Outcomes - 30 Day | All-Cause Mortality | 625 (1.3%) |
| | All-Cause Readmision or Mortality (ACread\|ACmort) | 6506 (13.1%) |
| | CVD Readmission or Mortality (CVread\|ACmort) | 3663 (7.4%) |
| | CVD Primary Dx Readmission or Mortality (PCVread\|ACmort) | 2655 (5.4%) |
| Outcomes - 6 Month | All-Cause Mortality | 2195 (4.4%) |
| | All-Cause Readmision or Mortality (ACread\|ACmort) | 15655 (31.6%) |
| | CVD Readmission or Mortality (CVread\|ACmort) | 9004 (18.2%) |
| | CVD Primary Dx Readmission or Mortality (PCVread\|ACmort) | 6854 (13.8%) |
| Top Primary Diagnosis | I48.9 - Atrial fibrillation and atrial flutter, unspecified | 4365 (8.7%) |
| | R07.4 - Chest pain, unspecified | 3786 (7.5%) |
| | I21.4 - Acute subendocardial myocardial infarction | 3786 (7.5%) |
| | I25.11 - Atherosclerotic heart disease, of native coronary artery | 3075 (6.1%) |
| | I50.0 - Congestive heart failure | 3042 (6.1%) |
| Top Primary Procedure | 38218-00 - Coronary angiography with left heart catheterisation | 3858 (10.5%) |
| | 38215-00 - Coronary angiography | 3554 (9.7%) |
| | 95550-03 - Allied health intervention, physiotherapy | 2700 (7.3%) |
| | 95550-09 - Allied health intervention, pharmacy | 2167 (5.9%) |
| | 38306-00 - Percutaneous insertion of 1 transluminal stent into single coronary artery | 2087 (5.7%) |

## Acute coronary syndrome

Table 2 shows the characteristics of encounters with a primary diagnosis of ACS. ACS was the largest cardiovascular subgroup in CardiacAI, accounting for a quarter of all encounters. The proportion of male patients (70.4%) was consistent with state-wide data[33]. However, the CardiacAI ACS cohort was younger than ACS patients in NSW generally with 30.1% of patients aged 75 years or over compared to 35% for all of NSW [33]. ACS patients had more admissions to ICU than the general CardiacAI population. ICU admission figures included both elective and unplanned and included patients going to ICU for post-cardiac surgery recovery. Unplanned readmission and mortality rates post-discharge were similar to rates reported in other studies[34]. The top five secondary discharge diagnoses were hypertension (50.5%), history of tobacco use (27.6%), current smoking (17.0%), and type 2 diabetes (15.4%).

*Table 2 Acute coronary syndrome encounter characteristics*

| **Measure** | | **Result** |
|---|---|---|
| Patients | | 10818 |
| Encounters | | 12518 |
| Sex | Male | 8814 (70.4%) |
| Age | Mean (SD) | 68.2 (12.4) |
| | Median [Min, Max] | 69.0 [19.0,103.0] |
| | <45 | 492 (3.9%) |
| | 45-54 | 1569 (12.5%) |
| | 55-64 | 2993 (23.9%) |
| | 65-74 | 3691 (29.5%) |
| | >=75 | 3773 (30.1%) |
| Indigenous Status | ABTSI | 334 (2.7%) |
| Country of Birth | Australia | 7201 (57.5%) |
| Language | NESB | 1598 (12.8%) |
| Length of Stay | Mean (SD) | 4.6 (6.1) |
| | Median [Min, Max] | 2.8 [0.0,143.6] |
| | <= 24 hrs | 2693 (21.5%) |
| | 1-3 days | 3979 (31.8%) |
| | 3-7 days | 3408 (27.2%) |
| | 7-14 days | 1671 (13.3%) |
| | >14 days | 767 (6.1%) |
| Admission Type | Emergency | 7811 (62.4%) |
| Outcomes - In-Hospital | ICU Admission | 1699 (13.6%) |
| | All-Cause Mortality | 238 (1.9%) |
| Outcomes - 30 Day | All-Cause Mortality | 105 (0.9%) |
| | All-Cause Readmision or Mortality (ACread\|ACmort) | 1508 (12.3%) |
| | CVD Readmission or Mortality (CVread\|ACmort) | 882 (7.2%) |
| | CVD Primary Dx Readmission or Mortality (PCVread\|ACmort) | 651 (5.3%) |
| Outcomes - 6 Month | All-Cause Mortality | 351 (2.9%) |
| | All-Cause Readmision or Mortality (ACread\|ACmort) | 3413 (27.8%) |
| | CVD Readmission or Mortality (CVread\|ACmort) | 1900 (15.5%) |
| | CVD Primary Dx Readmission or Mortality (PCVread\|ACmort) | 1475 (12.0%) |
| Top 5 Secondary Diagnosis | I25.11 - Atherosclerotic heart disease, of native coronary artery | 6584 (54.0%) |
| | U82.3 - Hypertension | 6159 (50.5%) |
| | Z86.43 - Personal history of tobacco use disorder | 3361 (27.6%) |
| | Z72.0 - Tobacco use, current | 2078 (17.0%) |
| | E11.9 - Type 2 diabetes mellitus without complication | 1879 (15.4%) |
| Top 5 Procedures | 38215-00 - Coronary angiography | 2529 (23.4%) |
| | 38218-00 - Coronary angiography with left heart catheterisation | 2526 (23.3%) |
| | 38306-00 - Percutaneous insertion of 1 transluminal stent into single coronary artery | 1981 (18.3%) |
| | 38500-00 - Coronary artery bypass, using 1 LIMA graft | 707 (6.5%) |
| | 38306-01 - Percutaneous insertion of 2 or more transluminal stents into single coronary artery | 573 (5.3%) |

## Heart Failure

Table 3 shows the characteristics of encounters with a primary diagnosis of HF. HF encounters made up a much smaller proportion of CardiacAI encounters (6.9%). These patients were more likely to be male (57.7%) had an older median age of 81 years and included a higher proportion of persons born outside of Australia (53.1%) compared with the other subgroups. Very few encounters were elective, with 92.5% being unplanned (emergency) encounters. The rate of in-hospital mortality was significantly lower (3.4% of encounters) than that of other Australian studies. The NSW and ACT HF Snapshot reported 6% in-hospital mortality in 2013,[35] and a study using linked administrative data found the in-hospital mortality rate dropped from 7.9% to 5.1% over the years 2010-2015[36]. This could indicate that patients are increasingly being transferred to a hospice or home for their end-of-life care, rather than dying in hospital. However, 30 day mortality in this study was 4.3% which is similar to the rate reported by Labrosciano et al[36] of 4.1% after discharge, which could suggest there have been improvements in the recognition and treatment of patients experiencing an acute heart failure episode. The 30 day ACread|ACmort rate was 24.5%, or 20.2% for all-cause readmission only, which is similar to the 17-27% rates reported by other Australian studies[34].

*Table 3 Heart failure encounter characteristics*

| Measure | | Result |
|---|---|---|
| Patients | | 2689 |
| Encounters | | 3484 |
| Sex | Male | 2010 (57.7%) |
| Age | Mean (SD) | 78.8 (11.3) |
| | Median [Min, Max] | 81.0 [21.0,103.0] |
| | <45 | 53 (1.5%) |
| | 45-54 | 101 (2.9%) |
| | 55-64 | 272 (7.8%) |
| | 65-74 | 635 (18.2%) |
| | >=75 | 2423 (69.5%) |
| Indigenous Status | ABTSI | 38 (1.1%) |
| Country of Birth | Australia | 1635 (46.9%) |
| Language | NESB | 885 (25.4%) |
| Length of Stay | Mean (SD) | 6.3 (7.6) |
| | Median [Min, Max] | 4.4 [0.0,293.5] |
| | <= 24 hrs | 204 (5.9%) |
| | 1-3 days | 854 (24.5%) |
| | 3-7 days | 1386 (39.8%) |
| | 7-14 days | 741 (21.3%) |
| | >14 days | 299 (8.6%) |
| Admission Type | Emergency | 3224 (92.5%) |
| Outcomes - In-Hospital | ICU Admission | 125 (3.6%) |
| | All-Cause Mortality | 120 (3.4%) |
| Outcomes - 30 Day | All-Cause Mortality | 143 (4.3%) |
| | All-Cause Readmision or Mortality (ACread\|ACmort) | 824 (24.5%) |
| | CVD Readmission or Mortality (CVread\|ACmort) | 634 (18.8%) |
| | CVD Primary Dx Readmission or Mortality (PCVread\|ACmort) | 485 (14.4%) |
| Outcomes - 6 Month | All-Cause Mortality | 517 (15.4%) |
| | All-Cause Readmision or Mortality (ACread\|ACmort) | 1855 (55.1%) |
| | CVD Readmission or Mortality (CVread\|ACmort) | 1489 (44.3%) |
| | CVD Primary Dx Readmission or Mortality (PCVread\|ACmort) | 1221 (36.3%) |
| Top 5 Secondary Diagnosis | U82.3 - Hypertension | 1800 (52.2%) |
| | U82.1 - Ischaemic heart disease | 1228 (35.6%) |
| | Z86.43 - Personal history of tobacco use disorder | 866 (25.1%) |
| | N17.9 - Acute kidney failure, unspecified | 843 (24.4%) |
| | I48.9 - Atrial fibrillation and atrial flutter, unspecified | 792 (23.0%) |
| Top 5 Procedures | 95550-03 - Allied health intervention, physiotherapy | 791 (29.7%) |
| | 95550-09 - Allied health intervention, pharmacy | 338 (12.7%) |
| | 92209-00 - Management of noninvasive ventilatory support, 24 hours or less | 178 (6.7%) |
| | 95550-01 - Allied health intervention, social work | 157 (5.9%) |
| | 38218-00 - Coronary angiography with left heart catheterisation | 148 (5.6%) |

### Atrial Fibrillation

Table 4 shows the characteristics of encounters with a primary diagnosis of AF. AF encounters made up 10.8% of all encounters with a similar proportion of males (55.7%), median age (72 years) and median LOS (1.4 days) to other AF hospitalisation studies[37,38]. Mortality was low at all follow up periods for the AF cohort (0.2% in-hospital, 0.7% at 30 days and 2.3% at 180 days). Unplanned readmission or mortality outcome rates for PCVread|ACmort were 7.9% and 18.8% at 30 and 180 days respectively. ACread|ACmort was 13% at 30 days and 31.5% at 6 months or 12.1% and 29% for all-cause readmission alone. Cardioversion was the most common primary procedure, occurring in 35.7% of all AF encounters and pacemakers were inserted in 7.2% of encounters. The all-cause 30 day readmission was higher than that reported by Woods et al. (9.9%)[38]. The rate of cardioversion was also higher than the rate reported by Weber et al. (29.5%)[39]. These differences may be due to differences in the cohort selection used by both studies which only included index AF hospitalisations, which could be associated with lower readmission and cardioversion rates[38,39]. The top secondary diagnoses included common cardiovascular risk factors and diseases. Interestingly, osteoarthritis was included in the top five secondary diagnoses, recorded in 12.1% of AF encounters.

*Table 4 Atrial fibrillation encounter characteristics*

| Measure | | Result |
|---|---|---|
| Patients | | 4297 |
| Encounters | | 5422 |
| Sex | Male | 3021 (55.7%) |
| Age | Mean (SD) | 70.0 (13.9) |
| | Median [Min, Max] | 72.0 [16.0,103.0] |
| | <45 | 302 (5.6%) |
| | 45-54 | 505 (9.3%) |
| | 55-64 | 973 (17.9%) |
| | 65-74 | 1505 (27.8%) |
| | >=75 | 2137 (39.4%) |
| Indigenous Status | ABTSI | 55 (1.0%) |
| Country of Birth | Australia | 3188 (58.8%) |
| Language | NESB | 756 (13.9%) |
| Length of Stay | Mean (SD) | 2.4 (2.9) |
| | Median [Min, Max] | 1.4 [0.0,50.2] |
| | <= 24 hrs | 1581 (29.2%) |
| | 1-3 days | 2541 (46.9%) |
| | 3-7 days | 951 (17.5%) |
| | 7-14 days | 302 (5.6%) |
| | >14 days | 47 (0.9%) |
| Admission Type | Emergency | 4241 (78.2%) |
| | ICU Admission | 27 (0.5%) |
| Outcomes - In-Hospital | All-Cause Mortality | 12 (0.2%) |
| Outcomes - 30 Day | All-Cause Mortality | 36 (0.7%) |
| | All-Cause Readmision or Mortality (ACread\|ACmort) | 701 (13.0%) |
| | CVD Readmission or Mortality (CVread\|ACmort) | 531 (9.8%) |
| | CVD Primary Dx Readmission or Mortality (PCVread\|ACmort) | 427 (7.9%) |
| Outcomes - 6 Month | All-Cause Mortality | 127 (2.3%) |
| | All-Cause Readmision or Mortality (ACread\|ACmort) | 1704 (31.5%) |
| | CVD Readmission or Mortality (CVread\|ACmort) | 1247 (23.0%) |
| | CVD Primary Dx Readmission or Mortality (PCVread\|ACmort) | 1018 (18.8%) |
| Top 5 Secondary Diagnosis | U82.3 - Hypertension | 2550 (52.3%) |
| | Z86.43 - Personal history of tobacco use disorder | 1012 (20.8%) |
| | U82.1 - Ischaemic heart disease | 921 (18.9%) |
| | U86.2 - Arthritis and osteoarthritis [primary] | 588 (12.1%) |
| | E11.9 - Type 2 diabetes mellitus without complication | 526 (10.8%) |
| Top 5 Procedures | 13400-00 - Cardioversion | 1276 (35.7%) |
| | 55118-00 - 2 dimensional real time transoesophageal ultrasound of heart | 573 (16.0%) |
| | 95550-09 - Allied health intervention, pharmacy | 382 (10.7%) |
| | 38353-00 - Insertion of subcutaneous cardiac pacemaker generator | 257 (7.2%) |
| | 95550-03 - Allied health intervention, physiotherapy | 206 (5.8%) |

## Syncope

Table 5 shows the characteristics of encounters with a primary diagnosis of syncope. Encounters where the primary diagnosis was syncope accounted for 3.4% of CardiacAI encounters. The proportion of males (57.7%) and median age (72 years) were similar to other studies[40,41]. Most encounters were classed as emergency (87.5%) and had a length of stay of less than 3 days (80.4%). There are limited Australian studies of syncope hospitalisation outcomes, despite patients hospitalised with syncope being at increased risk of cardiovascular events and mortality[42]. A recent Swedish study found that the average time from first syncope encounter to first cardiovascular event was 3.5 years[42]. However, findings from other international studies with a shorter follow up time (30 days) echo our readmission and mortality results[41,43]. CVread|ACmort and PCVread|ACmort rates were 2.9% and 2.2% at 30 days and 9.6% and 7.1% at 180 days. ACread|ACmort rates were 2.5 times that of the PCVread|ACmort rates for both 30 and 180 days. A loop recorder was

inserted in 15.3% of syncope encounters, and a pacemaker was inserted in 9.8% of syncope encounters.

*Table 5 Syncope encounter characteristics*

| Measure | | Result |
|---|---|---:|
| Patients | | 1632 |
| Encounters | | 1696 |
| Sex | Male | 979 (57.7%) |
| Age | Mean (SD) | 67.5 (19.1) |
| | Median [Min, Max] | 72.0 [15.0,106.0] |
| | <45 | 239 (14.1%) |
| | 45-54 | 148 (8.7%) |
| | 55-64 | 207 (12.2%) |
| | 65-74 | 385 (22.7%) |
| | >=75 | 717 (42.3%) |
| Indigenous Status | ABTSI | 10 (0.6%) |
| Country of Birth | Australia | 1085 (64.0%) |
| Language | NESB | 178 (10.5%) |
| Length of Stay | Mean (SD) | 2.2 (2.3) |
| | Median [Min, Max] | 1.5 [0.0,23.7] |
| | <= 24 hrs | 488 (28.8%) |
| | 1-3 days | 875 (51.6%) |
| | 3-7 days | 252 (14.9%) |
| | 7-14 days | 69 (4.1%) |
| | >14 days | 12 (0.7%) |
| Admission Type | Emergency | 1484 (87.5%) |
| Outcomes - In-Hospital | ICU Admission | 4 (0.2%) |
| | All-Cause Mortality | 2 (0.1%) |
| Outcomes - 30 Day | All-Cause Mortality | 10 (0.6%) |
| | All-Cause Readmision or Mortality (ACread\|ACmort) | 133 (7.8%) |
| | CVD Readmission or Mortality (CVread\|ACmort) | 49 (2.9%) |
| | CVD Primary Dx Readmission or Mortality (PCVread\|ACmort) | 37 (2.2%) |
| Outcomes - 6 Month | All-Cause Mortality | 40 (2.4%) |
| | All-Cause Readmision or Mortality (ACread\|ACmort) | 408 (24.1%) |
| | CVD Readmission or Mortality (CVread\|ACmort) | 163 (9.6%) |
| | CVD Primary Dx Readmission or Mortality (PCVread\|ACmort) | 120 (7.1%) |
| Top 5 Secondary Diagnosis | U82.3 - Hypertension | 692 (44.7%) |
| | U82.1 - Ischaemic heart disease | 275 (17.8%) |
| | Z86.43 - Personal history of tobacco use disorder | 247 (16.0%) |
| | U73.9 - Unspecified activity | 230 (14.9%) |
| | U86.2 - Arthritis and osteoarthritis [primary] | 180 (11.6%) |
| Top 5 Procedures | 95550-03 - Allied health intervention, physiotherapy | 141 (17.8%) |
| | 95550-09 - Allied health intervention, pharmacy | 122 (15.4%) |
| | 38285-00 - Insertion of subcutaneously implanted monitoring device | 121 (15.3%) |
| | 38353-00 - Insertion of subcutaneous cardiac pacemaker generator | 78 (9.8%) |
| | 95550-01 - Allied health intervention, social work | 34 (4.3%) |

## Discussion

In this study we have described the CardiacAI data repository, a valuable data resource for translational cardiovascular research and innovation. Through the creation of this research-ready, prospective data repository the CardiacAI project is poised to facilitate collaborative projects that examine contemporary health-related activity, characterise CVD cohorts, pragmatically monitor therapeutic interventions and patient outcomes, discover evidence-practice gaps and develop tools that address unmet clinical needs.

The CardiacAI project is supporting research projects which have already attracted cumulative funding of over $3M and include 1) development of a prototype hospital discharge risk stratification tool and clinical dashboard for cardiovascular patients, 2) expanding the current CardiacAI dataset to include stroke patients, remote monitoring data and two additional LHDs, and 3) developing infrastructure for next generation of clinical quality registries.

The robust CardiacAI infrastructure and governance framework is a key enabler. The CardiacAI-DGC provides a platform for data contributors and the project team to share control of data use. This also fosters a collaborative environment that encourages participation from local clinicians and researchers to answer local clinical questions and create tools that address local needs. The project also promotes participation from all health providers based in both the public and private sectors, by ensuring that data are held independently with input from all parties. Additionally, the ERICA platform's stringent data security controls provide assurance to data custodians and patients that data confidentiality is maintained.

The CardiacAI project has also facilitated easier access to detailed EMR data via the ERICA platform. This project's infrastructure and data repository not only provide a crucial resource for cardiovascular research but also for Australian health data science more generally. Utilisation of local data in the development of algorithms and data-driven tools, will potentially mitigate algorithmic biases that may arise from the use of tools developed exclusively on internationally sourced datasets[44]. This also ensures that the resultant algorithms perform as expected for Australian patient populations including indigenous and culturally and linguistically diverse communities (CALD) communities.

Finally, the inclusion of clinical text in the CardiacAI extraction provides a unique and enriched CVD dataset beyond traditional hospital administrative datasets. The CardiacAI project team are currently working on developing clinical text mining methods that extract important cardiovascular information from unstructured free-text. This work forms an important cornerstone of the CardiacAI project which will deliver critical data points for all projects and facilitate the creation of robust cardiovascular risk scores, phenotyping algorithms, clinical dashboards and other data tools[26].

## Current Limitations and Future Directions for CardiacAI

### New data sources

The CardiacAI Data Repository encompasses data from LHDs servicing a wide range of diverse populations, including those from CALD communities, indigenous peoples and individuals from various socioeconomic backgrounds[45,46] providing a wealth of data on patients living with CVD in NSW. Nonetheless, the repository does not yet incorporate stroke hospitalisations, which account for 11% of all CVD hospitalisations in Australia[8], or data from private hospitals which represent nearly half of all CVD hospitalisations[47].

Furthermore, CardiacAI currently lacks cardiac catheter procedure and diagnostic imaging data, and primary and community care datasets such as general practice and cardiac rehabilitation, which would provide valuable insights into CVD management during and after hospital discharge.

The CardiacAI project team is making significant progress towards expanding the repository to be more representative of the CVD population. The existing cohort is currently undergoing updates to incorporate stroke hospitalisations and emergency department deaths. Additionally, a retrospective EMR extract, based on cardiovascular diagnosis and procedure codes is being added, capturing admissions where patients have suffered a cardiovascular event but are admitted under non-cardiovascular specialists. These updates will ensure that the CardiacAI data includes all inpatients experiencing cardiac events.

The project is also expanding to two additional LHDs in NSW, one in regional NSW and the other in metropolitan Sydney, serving diverse CALD, indigenous and socioeconomic populations. This expansion will effectively double the size of the CardiacAI repository and address issues of diversity. EMR data for these new admission types and locations is expected in 2023. Moreover, several other LHDs and healthcare providers have shown interest in the project, and the team are seeking to iteratively incorporate new data sources either through sharing of the existing EMR extraction scripts or by adapting the extraction scripts to new hospital EMR systems.

The team are also progressing towards extracting data from other sources outside of hospitalisation data. Data from the TeleClinical Care (TCC)[48] telemonitoring application designed for patients who have experienced an ACS, heart failure or stroke event will be added to the repository In 2023. This innovative application is being rolled out to various hospitals in NSW, providing a previously unavailable data source of regular biometric observations, medication schedules and patient reported measures from patients while in the community. Additionally, the team is exploring the inclusion of cardiac catheter procedure, electrocardiogram and cardiac imaging data and community-based data sources such as pathology, radiology, cardiac rehabilitation services, heart failure outreach, and general practice (GP) data, either directly from GP EMR systems or through linkage with administrative data such as the Medicare Benefits Scheme (MBS) and Pharmaceutical Benefits Scheme (PBS).

### Data Quality and Standardisation

Undoubtably, EMR data is a valuable resource for the rapid generation of real-world evidence and the creation of tools to support clinical decision making. However, it is known to suffer from data quality issues[49,50]. While the CardiacAI dataset has undergone extensive testing to ensure it contains the correct information as recorded in the medical record, a detailed and systematic assessment of its quality in terms of correctness, completeness, concordance, and plausibility is underway and will be reported in the near future.
To facilitate the participation of CardiacAI data users in international studies, we also plan to extend our existing Australian-based data standardisation procedure to the Observational Medical Outcomes Partnership Common Data Model (OMOP CDM)[51], following previous work in this area[52,53]. The OMOP CDM is the lead data model used internationally to support real-world evidence and it is maintained and used by the Observational Health Data Science and Informatics (OHDSI) community[54].

Additionally, the team will explore the validation of structured information against concepts extracted from clinical text, as well as applying other established tools that offer more detailed EMR data quality assessment features[53,55]. By taking these steps, we aim to maximise the value of the CardiacAI dataset, ensuring the reliability and accuracy of its clinical information.

## Conclusion

The CardiacAI project has established a unique cardiovascular dataset that holds great potential for generating novel insights and innovations in the prevention of secondary cardiovascular events and the improvement of care and quality of life of those living with CVD. Leveraging its existing robust ethical, governance and data security infrastructure, the project team is committed to further expanding on its success, with the ultimate goal of becoming a comprehensive state-wide or national cardiovascular data asset.

## Acknowledgements

We extend our gratitude to SESLHS and ISLHD for their continued participation in the CardiacAI project. We also acknowledge the contribution of the Centre for Health Record Linkage in facilitating linkage of the EMR data with population health datasets.

## Ethical Approval

## Data Availability Statement

The data underlying this article cannot be shared publicly due to Australian privacy legislation. Researchers may apply to access the data within the CardiacAI platform by submitting an expression of interest to support@cardiacai.org. More information about data sharing in the CardiacAI project can be found at https://www.cardiacai.org/data. Study code is available at https://github.com/CardiacAI-AU

## Conflict of Interest

The authors declare no conflicts of interest.

## Funding

The Cardiac Analytics and Innovation project is a collaborative effort between the South Eastern Sydney Local Health District (SESLHD), Illawarra Shoalhaven Local Health District (ISLHD) and the University of New South Wales (UNSW) Centre for Big Data Research in Health (CBDRH), who have provided in-kind funding to the project. Other financial support has been provided by the 2020 Medical Research Future Fund (MRFF) Cardiovascular Health Mission Grant program, and the 2021 UNSW Medicine's Cardiac Vascular and Metabolic Medicine Big Ideas Seed Grant program. The views expressed in this article represent those of the authors and not the funding organisations.

# References

1. Shortreed SM, Cook AJ, Coley RY, Bobb JF, Nelson JC. Challenges and Opportunities for Using Big Health Care Data to Advance Medical Science and Public Health. *American Journal of Epidemiology* 2019;**188**:851-861. doi: 10.1093/aje/kwy292
2. Su J, Zhang Y, Ke Q-Q, Su J-K, Yang Q-H. Mobilizing artificial intelligence to cardiac telerehabilitation. *Reviews in Cardiovascular Medicine* 2022;**23**:045. doi: 10.31083/j.rcm2302045
3. Maddox TM, Albert NM, Borden WB*, et al.* The learning healthcare system and cardiovascular care: a scientific statement from the American Heart Association. *Circulation* 2017;**135**:e826-e857. doi: 10.1161/CIR.0000000000000480
4. Olsen L, Aisner D, McGinnis JM. The learning healthcare system: workshop summary. 2007. doi: 10.17226/11903
5. Zurynski Y, Smith CL, Vedovi A*, et al.* Mapping the learning health system: A scoping review of current evidence. In. *Australian Institute of Health Innovation and the NHMRC Partnership Centre for Health System Sustainability*: Australian Institute of Health Innovation, Macquarie University; 2020.
6. Bhavnani Sanjeev P, Parakh K, Atreja A*, et al.* 2017 Roadmap for Innovation—ACC Health Policy Statement on Healthcare Transformation in the Era of Digital Health, Big Data, and Precision Health. *Journal of the American College of Cardiology* 2017;**70**:2696-2718. doi: 10.1016/j.jacc.2017.10.018
7. Australian Institute of Health and Welfare. Health expenditure Australia 2016–17. https://www.aihw.gov.au/getmedia/e8d37b7d-2b52-4662-a85f-01eb176f6844/aihw-hwe-74.pdf.aspx?inline=true
8. Australian Institute of Health and Welfare. Heart, stroke and vascular disease: Australian facts. https://www.aihw.gov.au/reports/heart-stroke-vascular-diseases/cardiovascular-health-compendium
9. Australian Institute of Health and Welfare. Australian Burden of Disease Study 2022. https://www.aihw.gov.au/reports/burden-of-disease/australian-burden-of-disease-study-2022
10. Australian Health Ministers' Advisory Council. National Strategic Framework for Chronic Conditions: All Australians live healthier lives through effective prevention and management of chronic conditions. https://www.health.gov.au/resources/publications/national-strategic-framework-for-chronic-conditions
11. Johnson KW, Torres Soto J, Glicksberg Benjamin S*, et al.* Artificial Intelligence in Cardiology. *Journal of the American College of Cardiology* 2018;**71**:2668-2679. doi: 10.1016/j.jacc.2018.03.521
12. Kao DP, Trinkley KE, Lin C-T. Heart Failure Management Innovation Enabled by Electronic Health Records. *JACC: Heart Failure* 2020;**8**:223-233. doi: https://doi.org/10.1016/j.jchf.2019.09.008
13. Roca J, Tenyi A, Cano I. Paradigm changes for diagnosis: using big data for prediction. *Clinical Chemistry and Laboratory Medicine (CCLM)* 2019;**57**:317-327. doi: doi:10.1515/cclm-2018-0971

14. Srinivasan U, Rao S, Ramachandran D, Jonas D. Flying Blind: Australian Consumers and Digital Health. https://flyingblind.cmcrc.com/files/files/Flying-Blind--Australian-Consumers-and-Digital-Health.pdf (1)
15. Ehrenstein V, Kharrazi H, Lehmann H, Taylor CO. Obtaining data from electronic health records. In. *Tools and technologies for registry interoperability, registries for evaluating patient outcomes: A user's guide, 3rd edition, Addendum 2 [Internet]*: Agency for Healthcare Research and Quality (US); 2019.
16. Allen J, Adams C, Flack F. The role of data custodians in establishing and maintaining social licence for health research. *Bioethics* 2019;**33**:502-510. doi: 10.1111/bioe.12549
17. Denton N, Molloy M, Charleston S*, et al.* Data silos are undermining drug development and failing rare disease patients. *Orphanet Journal of Rare Diseases* 2021;**16**. doi: 10.1186/s13023-021-01806-4
18. Garmire LX, Gliske S, Nguyen QC*, et al.* The training of next generation data scientists in biomedicine. In: *Biocomputing 2017*. *2017*, p.640-645. World Scientific.
19. Chawinga WD, Zinn S. Global perspectives of research data sharing: A systematic literature review. *Library & Information Science Research* 2019;**41**:109-122. doi: https://doi.org/10.1016/j.lisr.2019.04.004
20. Cerner. Cerner Millennium. https://www.cerner.com/se/en/solutions/millennium (12th May 2021)
21. Penderson B. EMR in Australia: The state of the nation. https://www.hisa.org.au/slides/hic18/tue/BrucePedersen.pdf
22. NSW Health. Centre for Health Record Linkage (CHeReL). https://www.cherel.org.au/ (4th May 2021)
23. Desai T, Ritchie F, Welpton R. Five Safes: designing data access for research. https://core.ac.uk/download/pdf/323894811.pdf
24. University of New South Wales. ERICA - E-Research Institutional Cloud Architecture. https://research.unsw.edu.au/erica (4th May 2021)
25. Statistics ABo. Data Confidentiality Guide. https://www.abs.gov.au/about/data-services/data-confidentiality-guide (2022)
26. Percha B. Modern Clinical Text Mining: A Guide and Review. *Annual Review of Biomedical Data Science* 2021;**4**:165-187. doi: 10.1146/annurev-biodatasci-030421-030931
27. Liu L, Perez-Concha O, Nguyen A, Bennett V, Jorm L. De-identifying Australian hospital discharge summaries: An end-to-end framework using ensemble of deep learning models. *Journal of Biomedical Informatics* 2022;**135**:104215. doi: https://doi.org/10.1016/j.jbi.2022.104215
28. Joshy G, Korda R, Abhayaratna W, Soga K, Banks E. Categorising major cardiovascular disease hospitalisations from routinely collected data. *Public Health Research & Practice* 2015;**25**. doi: 10.17061/phrp2531532
29. Centre for Epidemiology and Evidence HN. NSW Health Cardiovascular Hospitalisations. https://www.healthstats.nsw.gov.au/#/r/106749 (15/02/2023)
30. Toner L, Koshy AN, Ko J, Driscoll A, Farouque O. Clinical characteristics and trends in heart failure hospitalizations: an Australian experience during the COVID-19 lockdown. *Heart Failure* 2020;**8**:872-875. doi: doi: 10.1016/j.jchf.2020.05.014
31. Sutherland K, Chessman J, Zhao J*, et al.* Impact of COVID-19 on healthcare activity in NSW, Australia. *Public health research & practice* 2020;**30**. doi: 10.17061/phrp3042030

32. Cho K, Femia G, Lee R, *et al.* Exploration of Cardiology Patient Hospital Presentations, Health Care Utilisation and Cardiovascular Risk Factors During the COVID-19 Pandemic. *Heart, Lung and Circulation* 2023. doi: https://doi.org/10.1016/j.hlc.2022.11.013
33. Centre for Epidemiology and Evidence HealthStats NSW. Coronary heart disease hospitalisations. https://www.healthstats.nsw.gov.au/#/r/106753 (15/02/2023)
34. Labrosciano C, Air T, Tavella R, Beltrame JF, Ranasinghe I. Readmissions following hospitalisations for cardiovascular disease: a scoping review of the Australian literature. *Australian Health Review* 2020;**44**:93. doi: 10.1071/ah18028
35. Newton PJ, Davidson PM, Reid CM, *et al.* Acute heart failure admissions in New South Wales and the Australian Capital Territory: the NSW HF Snapshot Study. *Med J Aust* 2016;**204**:113-113. doi: 10.5694/mja15.00801
36. Labrosciano C, Horton D, Air T, *et al.* Frequency, trends and institutional variation in 30‐day all‐cause mortality and unplanned readmissions following hospitalisation for heart failure in Australia and New Zealand. *European Journal of Heart Failure* 2021;**23**:31-40. doi: 10.1002/ejhf.2030
37. Weber C, Hung J, Hickling S, *et al.* Incidence, predictors and mortality risk of new heart failure in patients hospitalised with atrial fibrillation. *Heart* 2021;**107**:1320. doi: 10.1136/heartjnl-2020-318648
38. Woods T-J, Ngo L, Speck P, Kaambwa B, Ranasinghe I. Thirty-Day Unplanned Readmissions Following Hospitalisation for Atrial Fibrillation in Australia and New Zealand. *Heart, Lung and Circulation* 2022;**31**:944-953. doi: https://doi.org/10.1016/j.hlc.2022.02.006
39. Weber C, Hung J, Hickling S, *et al.* Emergent readmission and long-term mortality risk after incident atrial fibrillation hospitalisation. *Heart* 2023;**109**:380. doi: 10.1136/heartjnl-2022-321560
40. Simos P, Scott I. Appropriate use of transthoracic echocardiography in the investigation of general medicine patients presenting with syncope or presyncope. *Postgraduate Medical Journal* 2022:postgradmedj-2021-141416. doi: 10.1136/postgradmedj-2021-141416
41. Sandhu RK, Sheldon RS, Savu A, Kaul P. Nationwide Trends in Syncope Hospitalizations and Outcomes From 2004 to 2014. *Canadian Journal of Cardiology* 2017;**33**:456-462. doi: https://doi.org/10.1016/j.cjca.2016.11.005
42. Yasa E, Ricci F, Magnusson M, *et al.* Cardiovascular risk after hospitalisation for unexplained syncope and orthostatic hypotension. *Heart* 2018;**104**:487-493. doi: 10.1136/heartjnl-2017-311857
43. Abulhamayel A, Savu A, Sheldon RS, Kaul P, Sandhu RK. Geographical Differences in Comorbidity Burden and Outcomes in Adults With Syncope Hospitalizations in Canada. *Canadian Journal of Cardiology* 2018;**34**:937-940. doi: https://doi.org/10.1016/j.cjca.2018.04.011
44. Gianfrancesco MA, Tamang S, Yazdany J, Schmajuk G. Potential Biases in Machine Learning Algorithms Using Electronic Health Record Data. *JAMA Internal Medicine* 2018;**178**:1544. doi: 10.1001/jamainternmed.2018.3763
45. South Eastern Sydney Local Health Disctrict. Vulnerable and priority populations in South Eastern Local Health District: Analysis of ABS Census 2016. https://www.seslhd.health.nsw.gov.au/sites/default/files/groups/PICH/Priority%20Populations/Population%20Profile_vulnerable%20and%20priority%20populations_SESLHD_ABS%20Census%202016.pdf

46.	Illawarra Shoalhaven Local Health District. Health Care Services Plan 2020-2030. https://www.islhd.health.nsw.gov.au/sites/default/files/Health%20Plans/Final%20Health%20Care%20Services%20Plan%202020-2030.pdf
47.	Nghiem S, Afoakwah C, Scuffham P, Byrnes J. A baseline profile of the Queensland Cardiac Record Linkage Cohort (QCard) study. *BMC Cardiovascular Disorders* 2022;**22**. doi: 10.1186/s12872-022-02478-z
48.	Indraratna P, Biswas U, McVeigh J*, et al.* TeleClinical Care: A Randomised Control Trial of a Smartphone-Based Model of Care for Patients with Heart Failure or Acute Coronary Syndrome. *Heart, Lung and Circulation* 2021;**30**:S101. doi: https://doi.org/10.1016/j.hlc.2021.06.026
49.	Wei W-Q, Denny JC. Extracting research-quality phenotypes from electronic health records to support precision medicine. *Genome Medicine* 2015;**7**. doi: 10.1186/s13073-015-0166-y
50.	Weiskopf NG, Weng C. Methods and dimensions of electronic health record data quality assessment: enabling reuse for clinical research. *Journal of the American Medical Informatics Association* 2013;**20**:144-151. doi: 10.1136/amiajnl-2011-000681
51.	Observational Health Data Sciences and Informatics (OHDSI). OMOP Common Data Model. https://www.ohdsi.org/data-standardization/the-common-data-model/ (15/12/2021 2021)
52.	Quiroz JC, Chard T, Sa Z*, et al.* Extract, transform, load framework for the conversion of health databases to OMOP. *PLOS ONE* 2022;**17**:e0266911. doi: 10.1371/journal.pone.0266911
53.	Blacketer C, Defalco FJ, Ryan PB, Rijnbeek PR. Increasing trust in real-world evidence through evaluation of observational data quality. *Journal of the American Medical Informatics Association* 2021;**28**:2251-2257. doi: 10.1093/jamia/ocab132
54.	Observational Health Data Sciences and Informatics. Observational Health Data Sciences and Informatics. https://www.ohdsi.org/ (30/03/2023)
55.	Khare R, Utidjian LH, Razzaghi H*, et al.* Design and Refinement of a Data Quality Assessment Workflow for a Large Pediatric Research Network. *EGEMS (Wash DC)* 2019;**7**:36. doi: 10.5334/egems.294

# Supplementary Material

## E-Research Institutional Cloud Architecture (ERICA)

The E-Research Institutional Cloud Architecture (ERICA) platform was established by the University of New South Wales. ERICA provides a secure cloud (Amazon Web Services, AWS) computing environment for research using sensitive microdata. ERICA setup and configuration is highly automated, minimising the potential for human error. At all external access points, users authenticate themselves using a single set of login credentials (account name and password) plus mandatory multi-factor authentication code (using smartphone). Researchers use the ERICA environment through remote-access Windows virtual workstations, with anti-virus and malware detection systems. Researchers have no access from within their ERICA project spaces to any external web sites, email or other communication channels. All research data held in ERICA are encrypted both at-rest and in movement and secure protocols are used for all communications and data movement. Researchers can only import or export data via a controlled and audited gateway mechanism known as the Hub (external eHub and internal iHub applications) and all data movements are fully logged and subject to virus checking and full-copy audit trails. A designated Project Controller checks and approves all inbound and outbound files. All other file or data ingress and egress mechanisms, including clipboard, email, messenger services, printing services and internet access, are blocked. Project workspaces are isolated from each other, and no data can be transferred between them except via the Hub. All users are required to undertake training in ethics, privacy, information security, statistical disclosure control and use of ERICA. Regular on-site and off-site backups of data are made. All off-site backups and archival data are encrypted prior to being transferred to secure off-site storage. ERICA has completed the full eHealth NSW Privacy Security Assurance Framework (PSAF) review and eHealth NSW has formally advised that ERICA is designed with adequate security control to protect NSW Health data.

More information about ERICA can be found at https://research.unsw.edu.au/erica

### CardiacAI data description

SESLHD and ISLHD were previously one district and so share an EMR instance. Therefore, patients admitted to Prince of Wales Hospital, St George Hospital, The Sutherland Hospital or The Wollongong Hospital have one PatientID across all the sites. The EMR admissions (encounters) have an EncounterID attached which uniquely identifies each hospital admission. Personal identifiers are not transferred to the CardiacAI Data Repository. The PatientID and EncounterID are unique identifiers from the back-end source EMR database and are not searchable in front-end EMR applications. Patients that visit the emergency department only and are not admitted as an inpatient are not currently included in the CardiacAI Data Repository.

Data extracted from the EMR include both structured and unstructured data types. The EMR data elements are listed in Table 6. The full repository data dictionary can be found at [www.CardiacAI.org/data-dictionary](www.CardiacAI.org/data-dictionary). Pathology and vital sign measurements were selected based on expert clinical opinion of their relevance to the projects aims. A full list of the included measurements can be found at [www.CardiacAI.org/measurement-concepts](www.CardiacAI.org/measurement-concepts). The EMR data includes a comprehensive list of unstructured clinical documents created during the patient's admission. The full list of document types included in the repository can be found at [www.CardiacAI.org/document-concepts](www.CardiacAI.org/document-concepts).

*Table 6. CardiacAI Data Types*

| Data Type | Data Elements |
|---|---|
| Structured | Patient demographics |
| | Hospital encounter details |
| | Ward movements |
|  | Triage categories |
|  | Admission diagnoses codes (SNOMED) |
|  | Diagnoses and other conditions |
|  | Allergies |
|  | Surgical case details* |
|  | Surgical procedure codes (ACHI)* |
|  | Medication orders |
|  | Medication administrations |
|  | Medication history on admission |
|  | Medication orders on discharge |
|  | Medication reconciliation events |
|  | Blood product transfusions |
|  | Pathology test results |
|  | Microbiology test results |
|  | Medical Imaging test details** |
|  | Vital Signs |
| Unstructured | Triage reports |
| | Pre-admission forms |
| | Operation reports* |
| | Clinical Progress Notes |
| | Clinical deterioration/care escalation notes |
| | Mental health assessments |
| | Discharge Summaries |
| | Allied Health Notes |
| | Assessments and Care Plans |
| | ECG notes (imaging and signal data is not included yet) |

\* Surgical case data does not include cardiac catheterisation procedures.

\*\* Medical imaging data does not include actual images.

## Population Health Dataset Data Dictionaries

Data dictionaries for the linked population health data can be found through the following links:

*Table 7. Population Health Dataset Data Dictionaries*

| Population health dataset | Data dictionary link |
|---|---|
| Admitted Patient Data Collection (APDC) | https://www.cherel.org.au/media/38875/admitted-patient-data-collection_data-dictionary_aug_2022.docx |
| Emergency Department Data Collection (EDDC) | https://www.cherel.org.au/media/60057/emergency-department-data-collection-data-dictionary_aug_2022.docx |
| Registry of Births Deaths and Marriages (RBDM) & Cause of Death Unit Record File (COD-URF) | https://www.cherel.org.au/media/60080/nsw-mortality-data-dictionary_march_2020.docx |

## Top Other Diagnosis and Procedure Codes

Top primary diagnosis and procedure codes for non-ACS, HF, AF or Syncope admissions in the CardiacAI data repository.

*Table 8 Top five primary diagnosis and procedure codes for other encounter types*

|  |  | result |
|---|---|---|
| **Primary Diagnosis** | R07.4 - Chest pain, unspecified | 3781 (13.99%) |
|  | R07.3 - Other chest pain | 1081 (4.0%) |
|  | I70.21 - Atherosclerosis of arteries of extremities with intermittent claudication | 814 (3.01%) |
|  | Z45.0 - Adjustment and management of cardiac device | 797 (2.95%) |
|  | I35.0 - Aortic (valve) stenosis | 699 (2.59%) |
| **Primary Procedure** | 95550-03 - Allied health intervention, physiotherapy | 1324 (7.0%) |
|  | 35303-06 - Percutaneous transluminal balloon angioplasty | 1166 (6.17%) |
|  | 95550-09 - Allied health intervention, pharmacy | 1140 (6.03%) |
|  | 38353-00 - Insertion of subcutaneous cardiac pacemaker generator | 1095 (5.79%) |
|  | 38218-00 - Coronary angiography with left heart catheterisation | 1077 (5.7%) |